\documentclass[aps,prl,prabib,twocolumn,nopacs]{revtex4}
\usepackage{graphicx}
\usepackage{amsmath}
\usepackage{amssymb}
\usepackage{amsfonts}
\usepackage{bm}

\begin{document}

\title{The unitary three-body problem in a trap
}

\affiliation{Laboratoire Kastler Brossel, \'Ecole Normale
Sup\'erieure, 24 rue Lhomond, 75231 Paris Cedex 05, France}

\author{F\'elix Werner}
\affiliation{Laboratoire Kastler Brossel, \'Ecole Normale
Sup\'erieure, 24 rue Lhomond, 75231 Paris Cedex 05, France}

\author{Yvan Castin}
\affiliation{Laboratoire Kastler Brossel, \'Ecole Normale
Sup\'erieure, 24 rue Lhomond, 75231 Paris Cedex 05, France}

\begin{abstract}
We consider either 3 spinless bosons or 3 equal mass spin-1/2 fermions, interacting {\it via} a short range
potential of infinite scattering length and trapped in an isotropic
harmonic potential. For a zero-range model, we obtain analytically 
the exact spectrum and eigenfunctions: 
for fermions all the states are universal; 
for bosons there is a coexistence of decoupled universal 
and efimovian states. All the universal states, 
even the {\it bosonic} ones, have a tiny 3-body loss rate.
For a finite range model, we numerically find for bosons a coupling
between zero angular momentum universal and efimovian states;
the coupling is so weak that,
for realistic values of the interaction range,
these bosonic universal states remain long-lived and observable.
\end{abstract}

%\vspace{1cm}

%\pacs{03.75.Ss, 05.30.Jp}
% Degenerate Fermi gases, Boson systems

%\date{\today}

\maketitle

With a Feshbach resonance, it is now possible 
to produce a stable quantum gas of fermionic atoms
in the unitary limit, i.e. with an
interaction of negligible range and scattering length $a=\infty$
%\cite{Thomas,Jin,Salomon_1}.
\cite{stable_unitary_gas}.
The properties of this gas, including its superfluidity,
are under active experimental investigation
%\cite{Jin ``pair condensate'',Ketterle vortices,Grimm crossover,Grimm_gap,Salomon_crossover,
%Hulet closed channel, Thomas capa_calo}.
\cite{crossover_properties}.
They have the remarkable feature
of being universal, as was tested in particular for the zero temperature
equation of state of the gas
%\cite{Pandharipande,Giorgini}.
\cite{fixed_node}.
In contrast, experiments with Bose gases at a Feshbach resonance
suffer from high loss rates
%\cite{bosons ketterle, bosons wieman, bosons rempe, bosons grimm},
\cite{manips_bosons,grimm_efimov,Braaten_review},
and even the existence of a unitary Bose gas phase is a very open subject \cite{Braaten_bose_gas}.

In this context, fully understanding the few-body unitary problem is a crucial step.
In free space, the unitary 3-boson problem has an infinite number of 
weakly bound states, the so-called Efimov states \cite{Efimov1}.
In a trap, it has efimovian states \cite{Pethick,Koehler} but also universal states whose energy
depends only on the trapping frequency \cite{Pethick}.
Several experimental groups are currently trapping a few particles at a node of an optical lattice
\cite{Greiner reseau} and are controlling the interaction
strength {\sl via} a Feshbach resonance.
Results have already been obtained for two particles per lattice node
\cite{Esslinger reseau fermions}, a case that was solved analytically
\cite{2corps_wilkens}.
Anticipating experiments with 3 atoms per node, we derive in this Letter 
exact expressions for {\it all} universal and efimovian eigenstates of the 3-body problem
for bosons (generalizing \cite{Pethick} to a non-zero angular momentum)
and for equal mass fermions in a trap. We also show the long lifetime of the universal states
and their observability in a real experiment, extending to universal states
the numerical study of \cite{Koehler}.

If the effective range and the true range of the interaction 
potential are negligible as compared to the de~Broglie wavelength of the 
3 particles,
the interaction potential can be replaced by the Bethe-Peierls contact
conditions on the wavefunction $\psi$: it exists a function $A$ such that
\begin{equation}
\psi(\mathbf{r}_1,\mathbf{r}_2,\mathbf{r}_3) =
\left(\frac{1}{r_{ij}}-\frac{1}{a}\right)
A(\mathbf{R}_{ij},\mathbf{r}_{k})
+O(r_{ij})
\label{contact}
\end{equation}
in the limit $r_{ij}\equiv |\mathbf{r}_i-\mathbf{r}_j|\rightarrow 0$ 
taken for fixed positions of the other
particle $k$ and of the center of mass $\mathbf{R}_{ij}$ of $i$ and $j$.
In the unitary limit considered in this paper, $a=\infty$.
When all the $r_{ij}$ are non zero, the
wavefunction $\psi$ obeys 
the non-interacting Schr\"odinger equation
\begin{equation}
\sum_{i=1}^3
\left[ -\frac{\hbar^2}{2m} \Delta_{\mathbf{r}_i} +
\frac{1}{2}m\omega^2\, r_i^2 \right]
\psi = E \psi.
\end{equation}
$\omega$ is the oscillation frequency and $m$ the mass of an atom.

\begin{figure}
\includegraphics[width=\columnwidth,clip=]{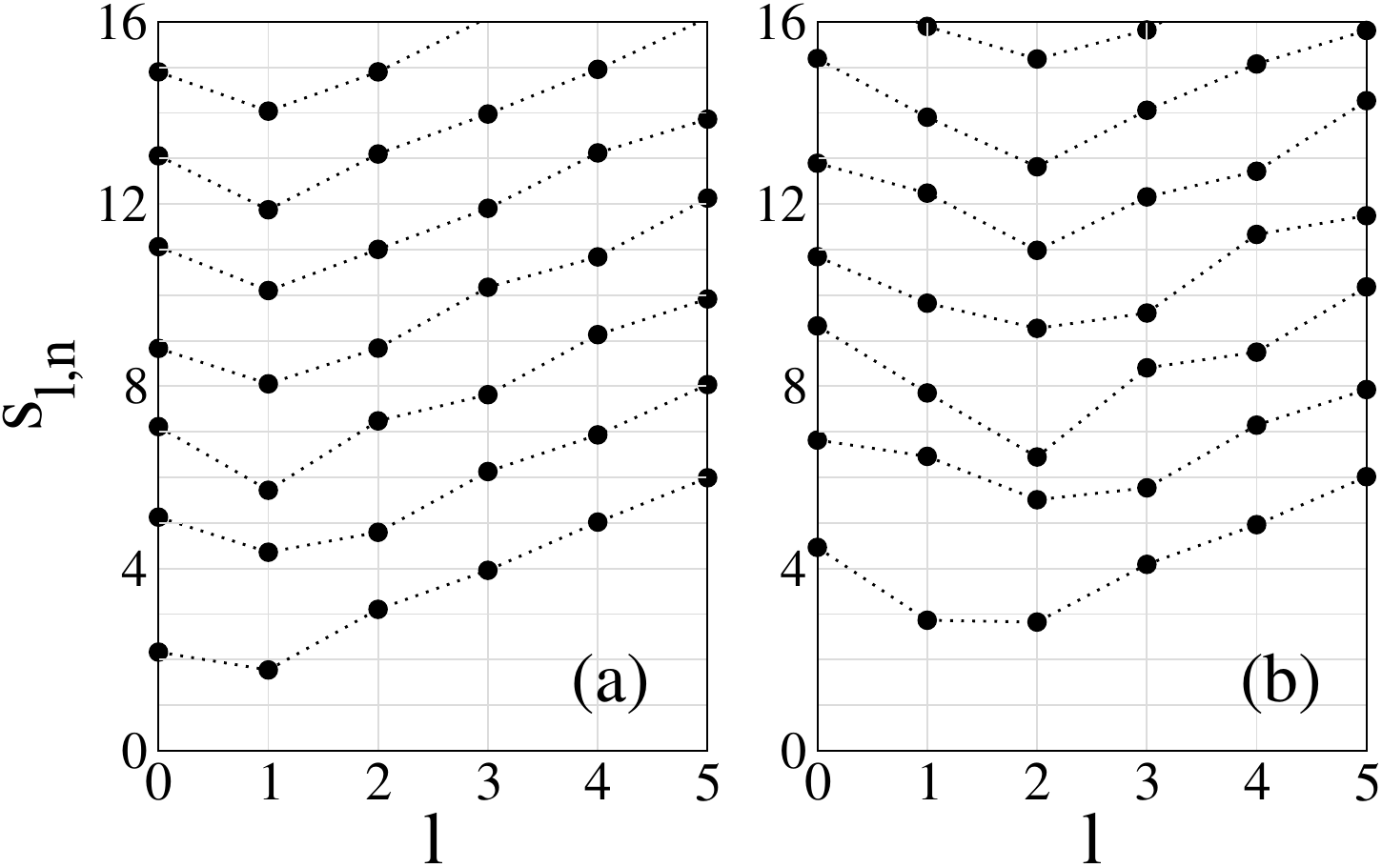}
\caption{The constants $s_{l,n}$ for (a) 3 equal mass fermions
and (b) 3 bosons, obtained by numerical solution of
the transcendental equation Eq.(\ref{eq:trans}).
We have not represented the $s_{l=0,n=0}$ solution for bosons, which is purely imaginary.
According to Eq.(\ref{eq:E}), each real $s_{l,n}$ gives rise to a semi-infinite ladder of
universal states. Note that the ground universal state has a total angular momentum $l=1$ for
fermions ($E \simeq 4.27 \hbar\omega$) and $l=2$ for bosons ($E \simeq 5.32 \hbar\omega$).}
\label{exposants}
\end{figure}

To solve this problem,
we extend the approach of Efimov \cite{Efimov1,Efimov2} to the trapped case, and obtain the form
\begin{equation}
\psi(\mathbf{r}_1,\mathbf{r}_2,\mathbf{r}_3)=\psi_{\rm cm}(\mathbf{C})
 F(R)
\left(1+\hat{Q}\right)
\frac{1}{r\rho} \varphi(\alpha)
Y^m_l\!(\mbox{\boldmath $\rho$}/{\rho}).
\end{equation}
Since the center of mass is separable for a harmonic trapping, we have singled out the wavefunction $\psi_{\rm cm}(\mathbf{C})$
of its stationary state of energy $E_{\rm cm}$, 
with $\mathbf{C}=(\mathbf{r}_1+\mathbf{r}_2+\mathbf{r}_3)/3$. The operator $\hat{Q}$ ensures the correct
exchange symmetry of $\psi$: for spinless bosons, $\hat{Q}=\hat{P}_{13}+\hat{P}_{23}$, where $\hat{P}_{ij}$ transposes particles $i$ and $j$;
for spin $1/2$ fermions, we assume a spin state $\uparrow\downarrow\uparrow$ so that $\hat{Q}=-\hat{P}_{13}$.
The Jacobi coordinates are $\mathbf{r}=\mathbf{r}_2-\mathbf{r}_1$ and 
$\mbox{\boldmath $\rho$}=(2\mathbf{r}_3-\mathbf{r}_1-\mathbf{r}_2)/\sqrt{3}$.
$Y_l^m$ is a spherical harmonic, $l$ being the total internal angular momentum of the system. The function $\varphi(\alpha)$, where
$\alpha=\arctan(r/\rho)$, solves the eigenvalue problem
\begin{eqnarray}
-\varphi''(\alpha) + \frac{l(l+1)}{\cos^2\alpha}\ \varphi(\alpha) &=& s^2\
 \varphi(\alpha)
\label{i}
\\
\varphi(\pi/2)&=&0
\label{ii}
\\
\varphi'(0) + \eta (-1)^l \frac{4}{\sqrt{3}}\ \varphi(\pi/3) &=& 0
\label{iii}
\end{eqnarray}
with $\eta=-1$ for fermions, $\eta=2$ for bosons. An analytical expression can be obtained for
$\varphi(\alpha)$ \cite{what_is_phi}, which leads to the transcendental equation for $s$ \cite{integer solutions}:
\begin{eqnarray}
\nonumber
& &
\Bigg[i^l\sum_{k=0}^l\frac{(-l)_k(l+1)_k}{k!}\frac{(1\!-\!s)_l}{(1\!-\!s)_k}
\Big(
    2^{-k}i(k-s)e^{is\frac{\pi}{2}}
\\
& &
+\eta (-1)^l\frac{4}{\sqrt{3}}e^{i\frac{\pi}{6}(2k+s)}
\Big) \Bigg] - [i \leftrightarrow -i]  = 0,
\label{eq:trans}
\end{eqnarray}
with the notation $(x)_n\equiv x (x+1)\ldots (x+n)$.
This equation is readily solved numerically: for each $l$, the solutions form
an infinite sequence $(s_{l,n})_{n\geq 0}$, see Fig.\ref{exposants}. 
As we show below, all solutions are real, except for bosons in the $l=0$ channel, where 
a single purely imaginary solution exists, $s_{l=0,n=0}\equiv s_0\simeq i\times
1.00624$, the well known Efimov solution.
Finally, the function $F(R)$, where the hyperradius is $R=\sqrt{(r^2+\rho^2)/2}$, solves the problem:
\begin{equation}
\left[-\frac{\hbar^2}{2m}\left(\frac{d^2}{dR^2}+\frac{1}{R}\frac{d}{dR}\right)+
U(R)\right] F(R)= (E-E_{\rm cm}) F(R)
\label{eq:sle}
\end{equation}
where $U(R)=\hbar^2 s^2/(2m R^2) + m\omega^2 R^2/2$, $s$ being one of the $s_{l,n}$.
This is the Schr\"odinger equation for a fictious particle of zero angular momentum
moving in two dimensions
in the potential $U(R)$.

When $s^2>0$, one takes $s>0$ and the solution is
\begin{equation}
F(R)= R^{s} e^{-R^2/2a_{\rm ho}^2}
L_q^{(s)}\!\left(R^2/a_{\rm ho}^2\right)
\label{wavefunctions}
\end{equation}
where $a_{\rm ho}=(\hbar/m\omega)^{1/2}$ is the harmonic oscillator length,
$L_q^{(\cdot)}$ is the generalized Laguerre polynomial of
degree $q$, $q$ being an arbitrary non-negative integer. 
The resulting spectrum for the 3-body problem is
\begin{equation}
E=E_{\rm cm} + (s_{l,n}+1+2q)\hbar\omega.
\label{eq:E}
\end{equation}
The quantum number $q$ leads to a semi-infinite ladder structure of the spectrum with a regular
spacing $2\hbar\omega$. This is related to the existence of a scaling solution
for the trapped unitary gas \cite{CRAS} and the subsequent embedding of the Hamiltonian
in a $SO(2,1)$ algebra \cite{Pitaevskii}, leading to an exact mapping between
trapped and free space universal states \cite{Tan_et_nous}.

When $s^2<0$, as is the case in the $l=n=0$ channel for bosons, the Schr\"odinger equation
Eq.(\ref{eq:sle}) does not define by itself an Hermitian problem and has to be supplemented by
a boundary condition for $R\rightarrow 0$  \cite{Morse,Danilov}:
\begin{equation}
F(R) \propto \mbox{Im}\, \left[\left(\frac{R}{R_t}\right)^{s_0}\right],
\end{equation}
where $R_t$ is an additional 3-body parameter.
For the resulting efimovian states, the function $F$ is given by
\begin{equation}
F(R) = R^{-1}\, W_{(E-E_{\rm cm})/2\hbar\omega,s_0/2}(R^2/a_{\rm ho}^2)
\end{equation}
where $W$ is a Whittaker function, and the energy solves:
\begin{eqnarray}
\mbox{arg}\, \Gamma\left[\frac{1+s_0-(E-E_{\rm cm})/\hbar\omega}{2}\right] &=&
-|s_0| \ln (R_t/a_{\rm ho})\nonumber \\
 +\mbox{arg}\, \Gamma(1+s_0) \ \mbox{mod}\ \pi. &&
\end{eqnarray}

We did not yet obtain all the 3-body eigenstates \cite{pas_tout}.
Indeed, all the above states satisfy the contact condition (\ref{contact}) with a {\textit{non-zero}} function $A$.
But there are wavefunctions of the unitary gas which {\textit{vanish}} when two
particles are at the same point; these are also eigenstates of the non-interacting case.
An example is the Laughlin state of the Fractional Quantum Hall Effect \cite{Laughlin}:
\begin{equation}
\psi=e^{-\sum_{i=1}^{3} r_i^2/2a_{\rm ho}^2}\prod_{1\leq n< m \leq 3}
\left[(x_n+i y_n)-(x_{m}+i y_{m})\right]^{|\eta|}.
\end{equation}
In the limit of high energies $E\gg\hbar\omega$, there are actually many of these $A\equiv 0$ states:
their density of states (DOS) is almost as high as the DOS of the non-interacting case:
\begin{equation}
\frac{\rho_{A\equiv 0}(E)}{\rho_{\rm non-inter}(E)} \underset{E\rightarrow \infty}{=} 1-
O\bigg(\Big(\frac{\hbar\omega}{E}\Big)^2\bigg).
\label{A=0}
\end{equation}
In contrast, the DOS of the $A\neq 0$ states is only
\begin{equation}
\frac{\rho_{A\neq 0}(E)}{\rho_{\rm non-inter}(E)} \underset{E\rightarrow \infty}{=} O\bigg(\Big(\frac{\hbar\omega}{E}\Big)^2\bigg).
\label{A neq 0}
\end{equation}
Eq.(\ref{A neq 0}) is a consequence of Eq.(\ref{almost integer}) given below.
We found Eq.(\ref{A=0}) by applying the rank theorem to the operator
$\psi_0(\mathbf{r}_1,\mathbf{r}_2,\mathbf{r}_3) \mapsto
\big( \psi_0(\mathbf{r}_1,\mathbf{r}_1,\mathbf{r}_3),$
$\psi_0(\mathbf{r}_1,\mathbf{r}_2,\mathbf{r}_1),
\psi_0(\mathbf{r}_1,\mathbf{r}_2,\mathbf{r}_{2})
\big)$
which associates, to each non-interacting eigenstate $\psi_0$ of energy $E$, $3$ functions of $2$ atomic positions,
and whose kernel is the space of $A\equiv 0$ states of energy $E$
\cite{rank details}.

This completes our derivation of {\sl all} eigenstates of the unitary 3-body problem in a trap.
Three types of states are obtained in general: universal eigenstates common to the non-interacting case,
universal interacting states, and efimovian states depending on a 3-body parameter.

We now prove that the Efimov effect is absent for 3 equal mass fermions. 
This fact is known but to our knowledge
not demonstrated.
Numerically one can only check the absence of imaginary solution of the
transcendental equation in some finite interval of $s$ and $l$.
Here we prove that for any $l$ and any imaginary $s$, there is no solution
to the problem (\ref{i},\ref{ii},\ref{iii}).
Let us assume that $s^2\leq l(l+1)$, and that (\ref{i},\ref{ii})
are satisfied. 
We will show that the quantity
$Q(l,s^2)\equiv\varphi'(0)-(-1)^l (4/\sqrt{3})\varphi(\pi/3)$
is non zero, which is incompatible with (\ref{iii}).
We rewrite (\ref{i}) as $\varphi''(\alpha)=u(\alpha)\varphi(\alpha)$.
This is Newton's equation, $\alpha$ being the time and $\varphi$ the position
of a fictitious particle subject to an {\it expelling} harmonic force 
with time dependent spring constant 
$u(\alpha;l,s^2)=\frac{l(l+1)}{\cos^2\alpha}-s^2 \geq 0$.
Eq.(\ref{ii}) imposes that this particle reaches the origin at `time' $\pi/2$.
The particle then should not reach the origin earlier, otherwise the
expelling force would prevent it from turning back to $\varphi=0$.
We thus can take the normalization $\varphi(0)=1$, which implies
$\varphi'(0)<0$ and $\varphi(\alpha)>0$ for $0\leq \alpha < \pi/2$.
Thus, $Q(l,s^2)<0$ for $l$ even.
For $l$ odd, one needs two intermediate results:
(i) $Q(l=1,s^2=2)<0$ (which we
check by explicit calculation); (ii) if $\varphi_1,\varphi_2$ are two solutions
with $u_2\geq u_1$, then $\varphi_2\leq\varphi_1$, and $Q_2\leq Q_1$:
because the spring constant for particle 2 is larger, particle 2 has to
start faster and
walk constantly ahead of particle 1 
in the race towards the origin, to satisfy Eq.(\ref{ii}).
Now the assumption $s^2\leq l(l+1)$ implies $u(\alpha;l,s^2)\geq u(\alpha;l=1,s^2=2)$. One concludes that:
$Q(l,s^2) \leq Q(l=1,s^2=2) < 0$.
For bosons, we proved similarly that all the $s^2$ are positive, except for
the well known $s_{n=0,l=0}\simeq i\times 1.00624$.

It appears clearly on Fig.\ref{exposants} that $s_{l,n}$ gets close
to an integer value $\bar{s}_{l,n}$ as soon as $l$ or $n$ increases, with
\begin{equation}
\begin{array}{lc}
\bar{s}_{l,n}=l+1+2n & \textrm{\, for\, } l\geq |\eta|\\
\bar{s}_{l,n}= 2n-l +(2\eta+11)/3 &  \textrm{\, for\, } l < |\eta| 
\end{array}.
\label{almost integer}
\end{equation}
To check this analytically, the transcendental equation is not useful. 
We rather applied semi-classical WKB techniques to the problem (\ref{i},\ref{ii},\ref{iii}), and obtained \cite{18}:  
\begin{eqnarray}
\label{eq:18}
\!\!\!\!\!\!\!\!\!\!\!\!s_{l,0}-\bar{s}_{l,0} &\underset{l\rightarrow\infty}{\sim}& \eta (-1)^{l+1}
2^{1-l} \big/ \sqrt{3\pi l} \\
\!\!\!\!\!\!\!\!\!\!\!\!s_{l,n}-\bar{s}_{l,n} &\underset{n\rightarrow\infty}{\sim}& \eta
\cos\left[\frac{\pi}{3}(l+1-n)\right]
\frac{(-1)^{l+n+1}4}{\pi\sqrt{3}\ n} 
\end{eqnarray}
\vskip -0.3cm
\begin{equation}
\underset{n}{\mathrm{max}}\ |s_{l,n}-\bar{s}_{l,n}|
\underset{l\rightarrow\infty}{\sim} |\eta| \frac{4\,
Ai_{\rm max}}{3^{7/12}\,\pi^{1/2}}\ \,l^{-5/6}
\end{equation}
with $Ai_{\rm max}\simeq0.5357$ the maximum of the Airy function.

We now discuss the lifetime of the 3-body states
found here in the trap,
due to 3-body recombination to a deeply bound molecular state.
The recombination rate is commonly estimated as
$\Gamma_{\rm loss} \propto  P \hbar/(m \sigma^2)$,
where $\sigma$ is the range of
the interaction potential, and $P$ is the
probability that $R<\sigma$
\cite{recette}.
Evaluating $P$ from the 3-body wavefunctions obtained above 
for the zero range model, this gives for $E$ not much larger than 
$\hbar\omega$:
\begin{equation}
\Gamma_{\rm loss}^{\rm univ} \propto \omega \left(\frac{\sigma}{a_{\rm ho}}\right)^{2s}
\label{Gamma_PP}
\end{equation}
for a universal state with exponent $s$,
and $\Gamma_{\rm loss}^{\rm efim} \propto \omega$
for an efimovian state.
Since $s\geq1.77$ for fermions and $s\geq2.82$ for bosons (Fig.\ref{exposants}),
Eq.(\ref{Gamma_PP}) indicates that the lifetime of universal states is
$\gg 1/\omega$ for $\sigma\ll a_{\rm ho}$.

\begin{figure}
\includegraphics[width=0.975\columnwidth,clip=]{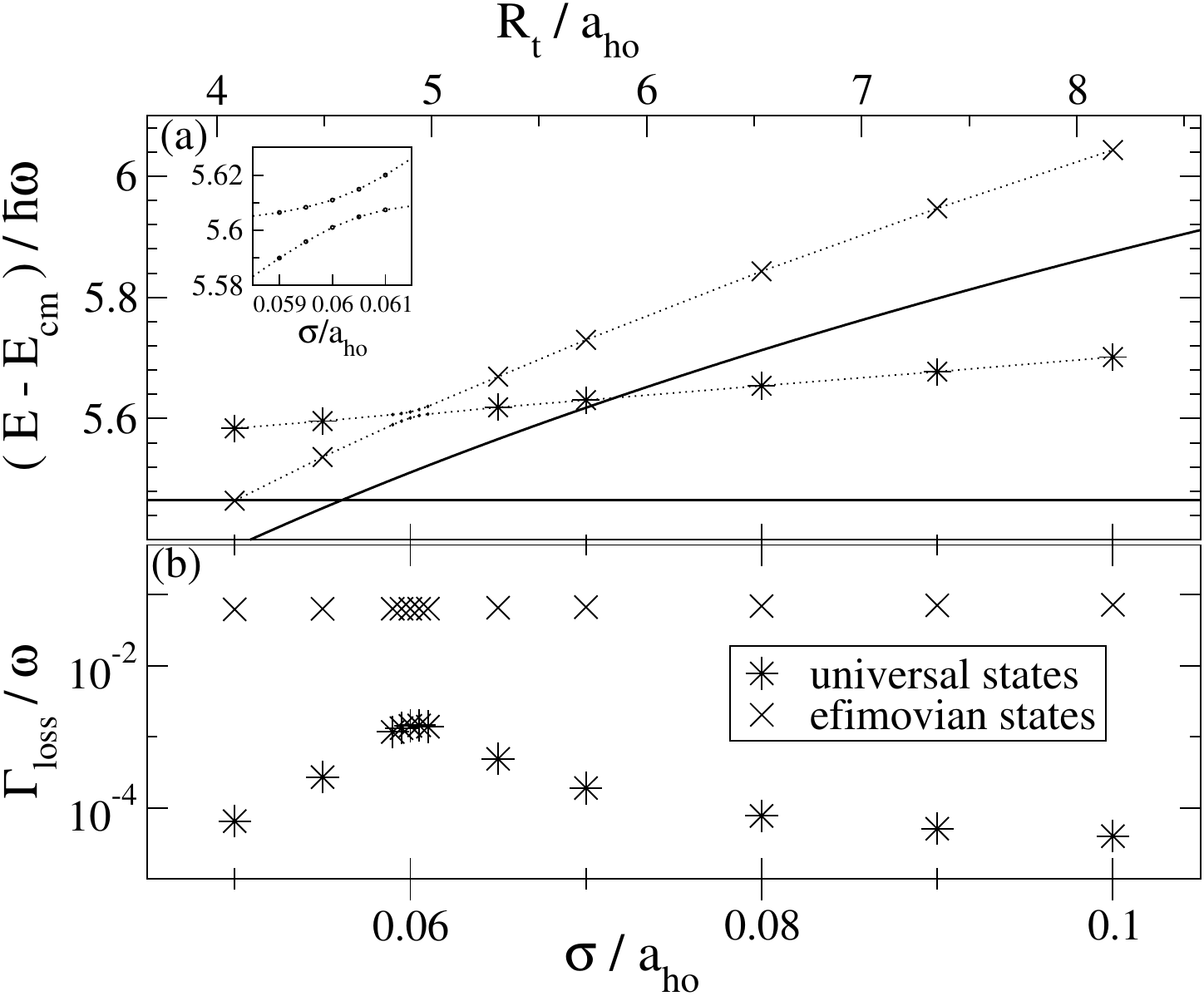}
\caption{Numerical solution of the separable potential model: 
(a) 3-body eigenenergies and (b) predicted 3-body loss rates (for 
the case of ${}^{133}$Cs, see text),
as a function of the potential range $\sigma$ (lower axis) and the 3-body
parameter $R_t$ (upper axis) \cite{what_is_R_t}.
(a) The lowest energy universal branch (stars) and an efimovian branch
(crosses) have a very weak avoided crossing (inset). The analytical predictions of the zero-range model 
(solid lines) are in good agreement with the numerics (except for the avoided crossing); a linear extrapolation of
the stars to $\sigma=0$ matches the zero-range result at the $10^{-3}$ level.
(b) The universal states have a loss rate much smaller than $\omega$.
}
\label{fig:sep}
\end{figure}

The existence of long-lived bosonic states is an unexpected feature that we now
investigate in a more realistic way.
The unitary three-body
problem in an isotropic harmonic trap may be realized experimentally
by trapping 3 atoms at a site of a deep optical lattice, and using
a Feshbach resonance.
For a broad Feshbach resonance, the effective range is of the order of the 
Van der Waals length, which is roughly one order of magnitude smaller than
$a_{\rm ho}$, for an usual lattice spacing of $\sim 0.5 \mu$m and a lattice depth of $\sim 50$ recoil energies.
This experimental situation is not deeply in the asymptotic regime
of a zero range potential. Moreover, in the zero range model, there are energy crossings between 
universal and efimovian states as a function of $R_t/a_{\rm ho}$ (see solid lines
in Fig.\ref{fig:sep}a); as we shall see, for a finite range, 
there is a coupling between $l=0$ universal and efimovian states, leading to avoided
crossings \cite{Ludovic}, and to an additional contribution to the loss rate of 
$l=0$ universal states not included in Eq.(\ref{Gamma_PP}).

We therefore solve a finite interaction range model, the Gaussian separable potential 
of range $\sigma$ \cite{Koehler}, defined as
\begin{equation}
\langle \mathbf{r}_1,\mathbf{r}_2 | V | \mathbf{r}_1', \mathbf{r}_2'\rangle =
-\frac{\hbar^2}{2\pi^{3/2}m\sigma^5} e^{-(r_{12}^2+r_{12}^{'2})/2\sigma^2}
\delta(\mathbf{R}_{12}-\mathbf{R}_{12}').
\end{equation}
This leads to an integral equation that we solve numerically.
In Fig.\ref{fig:sep}a, we show two $l=0$ energy branches as a function of $\sigma$, corresponding in the zero-range
model to the lowest $l=0$ universal state and to an efimovian branch. 
The smallness of the avoided crossing between the two branches shows that
the coupling due to the finite range of the interaction is weak: the energy splitting at the
avoided crossing is $\hbar \Omega\simeq0.01\hbar\omega$, see inset of Fig.\ref{fig:sep}a. 

We now revisit the calculation of the 3-body loss rate for bosons, since
Eq.(\ref{Gamma_PP}) neglects the contamination of the universal state by
the efimovian state.
To account for the losses we add to the Hamiltonian $H_{\rm sep}$ of the
separable potential model an antihermitian part leading to
the effective Hamiltonian in second quantized form
\begin{equation}
H_{\rm eff} = H_{\rm sep}-i B_3\,
\frac{\hbar^2\sigma^4}{12m}\int[\psi^\dag(\vec{r}\,)]^3[\psi(\vec{r}\,)]^3\,d\vec{r},
\end{equation}
where $B_3$ is a numerical factor, whose actual value
depends on short-range atomic and molecular physics.
Specializing to $^{133}{\rm Cs}$, 
we adjust the parameters of our model to $B_3=25$ and $\sigma=6.5{\rm nm}$ 
in order to reproduce the three-body loss rate measured in
a non-condensed gas for several negative values of $a$ in \cite{grimm_efimov}.
To obtain the loss rates shown in Fig.\ref{fig:sep}b,
we restricted $H_{\rm eff}$ to the two branches of Fig.\ref{fig:sep}a:  the eigenvalues of the resulting 2$\times$2 matrix
have complex parts $-i\hbar\Gamma_{\rm loss}/2$.
For the efimovian states, $\Gamma_{\rm loss}\simeq 0.07\omega$. For the
universal states $\Gamma_{\rm loss}$ 
is several orders of magnitude smaller;
this remains true on the avoided crossing, because the coupling $\Omega/2$ of the 
universal state to the efimovian state is much smaller
than the decay rate of the efimovian state \cite{Cohen}.

Experimentally,
if one starts with the non-interacting ground state,
a superposition of 3-body unitary eigenstates can be prepared
by switching suddenly the scattering length from zero to infinity.
The Bohr frequencies in the subsequent evolution of an observable would
give information on the 3-body spectrum.
For bosons, there will be a finite fraction of the sites where the three
atoms have a long lifetime. This fraction is equal to the probability
of having populated a universal state, which we calculate to be 
$\simeq 0.174$, a value dominated
by the contribution ($\simeq 0.105$) of the lowest $l=0$ universal state.

In summary, we obtained the complete analytical solution of
a zero-range unitary 3-body problem in a trap.
For bosons, there are efimovian and universal states,
while for equal mass fermions we proved that all states are universal.
All universal states are stable in the zero-range limit 
with respect to 3-body losses,
not only for fermions, but also for bosons.
From the numerical solution of a finite range model, we find that,
although the bosonic universal states of zero angular momentum
slightly mix with the efimovian states,
their lifetime remains much
larger than the oscillation period in the trap.

\begin{acknowledgments}
We thank L.\ Pricoupenko, D.\ Petrov,
T. K\"{o}hler, A. Bulgac, D. Bauer for very useful discussions;
T. Kr\"{a}mer {\it et al.} for their data.  LKB is a Unit\'e de
Recherche de l'ENS et de l'Universit\'e Paris
6, associ\'ee au CNRS. Our research group is a member of IFRAF.
\end{acknowledgments}


\begin{thebibliography}{99}
\bibitem{integer solutions}
The integer solutions must be eliminated 
($l=0,s=2$ and $l=1,s=1$ for fermions; $l=0,s=4$ and $l=1,s=3$ for bosons).
\bibitem{pas_tout}
The above solution is based on a sequence of ansatz, which misses 
some of the eigenstates.
\bibitem{rank details}
The rank theorem is: for an operator $T$ acting on a finite dimensional space ${\cal E}$,
${\rm dim}~{\cal E}~=~{\rm dim}({\rm Ker}\ T)+{\rm dim}({\rm Im}\ T)$.
An upper bound on ${\rm dim}({\rm Im}\ T)$ then leads to the result.
\bibitem{18} 
For (\ref{eq:18}) we used $\partial_{s^2} \varphi'(0)=\int_0^{\pi/2}
\varphi^2(\alpha)$
where $\varphi$ solves (\ref{i},\ref{ii}) with the normalization 
$\varphi(0)=1$.
\bibitem{what_is_R_t} We find
$\ln(R_t/\sigma)\simeq 4.40 \mod \pi/|s_0|$ for the separable potential
from the free space Efimov states energies.
\end{thebibliography}
\end{document}